\documentclass[a4paper,11pt]{article}

\usepackage[a4paper]{geometry}

\usepackage{amsmath}
\usepackage{amsfonts}
\usepackage{graphicx}
\usepackage{authblk}
\usepackage[font=small,labelfont=bf]{caption}
\usepackage[caption=false,font=footnotesize]{subfig}
\usepackage[colorlinks=true, linkcolor=blue, urlcolor=blue, citecolor=red, anchorcolor=green]{hyperref}
\usepackage[numbers,sort&compress]{natbib}
\usepackage{doi}

\providecommand{\keywords}[1]{\textbf{\small \textit{keywords:}} {\small #1}}

\title{Entanglement Distribution in Optical Networks}

\author[1]{Alex Ciurana}
\author[1]{Vicente Martin}
\author[1]{Jesus Martinez-Mateo}
\author[2]{Bernhard Schrenk}
\author[2]{Momtchil Peev}
\author[2]{Andreas Poppe}
\affil[1]{\small Facultad de Inform\'{a}tica, Universidad Polit\'{e}cnica de Madrid, \authorcr
Campus de Montegancedo, 28660 Boadilla del Monte, Madrid, Spain}
\affil[2]{\small Safety \& Security Department, AIT Austrian Institute of Technology GmbH, Donau-City-Strasse 1, 1220 Vienna, Austria}

\begin{document}

\maketitle

\begin{abstract}
The ability to generate entangled photon-pairs over a broad wavelength range opens the door to the simultaneous distribution of entanglement to multiple users in a network by using centralized sources and flexible wavelength-division multiplexing schemes. Here we show the design of a metropolitan optical network consisting of tree-type access networks whereby entangled photon-pairs are distributed to any pair of users, independent of their location. The network is constructed employing commercial off-the-shelf components and uses the existing infrastructure, which allows for moderate deployment costs. We further develop a channel plan and a network-architecture design to provide a direct optical path between any pair of users, thus allowing  classical and one-way quantum  communication as well as entanglement distribution. This allows the simultaneous operation of multiple quantum information technologies. Finally, we present a more flexible backbone architecture that pushes away the load limitations of the original network design by extending its reach, number of users and capabilities.
\end{abstract}


\section{Introduction}

Entanglement is one of the most striking features of the quantum world. This property, with no classical counterpart, can be used in many applications of quantum information processing, such as the well-known quantum key distribution (QKD)~\cite{Bennett_92}, superdense coding~\cite{Bennett_92b} or quantum teleportation~\cite{Bennett_93}, among others. All these applications require quantum and classical communications, thus needing two communication channels: a quantum channel and a conventional one. The former, governed by the laws of quantum mechanics, is used to exchange quantum states (typically photons) carrying quantum information while the latter is assumed to be an error-free channel. Further requirements can be imposed on the classical channel for certain purposes (e.g., it must be authentic if it is used to extract a secret key in QKD post-processing).

These applications of quantum information processing, in general, and entanglement, in particular, require point-to-point communications as well as a high degree of isolation from the environment to avoid decoherence and to enable noise-free quantum channel operation to allow detecting the weak quantum signals. Therefore, the simplest way to design a network supporting quantum channels is to use a point-to-point topology with dedicated links (typically optical fibers). Unfortunately, this is a very expensive and therefor, impractical solution. Using existing infrastructure and sharing it with other users of the communication channels (quantum or conventional ones) would reduce the deployment and operational costs to an acceptable level.

For this reason, the study of the integration of quantum signals in point-to-point links via multiplexing~\cite{Townsend_97b, Peters_09, Eraerds_10, Qi_10, Patel_14} or using dark fibers in the field~\cite{Hughes_96, Holloway_11, Dynes_12, Shimizu_14} has been a long standing topic in quantum communications. As a next logical step,  a point-to-multipoint network scenario such as the one provided by access networks in standard optical communication networks should be considered. In the non-entangled case, there have been proposals for access networks based on time division multiplexing (TDM) technologies \cite{Lancho_10, Capmany_10, Choi_11, Razavi_12, Frohlich_13, Aleksic_13, Martinez_14} and wavelength division multiplexing (WDM) \cite{Choi_10, Aleksic_13}. For entangled pairs of photons, only WDM-based access networks have been studied~\cite{Brassard_04, Lim_10, Ghalbouni_13, Herbauts_13}. Beyond these approaches, the next step has been to devise the integration of quantum communication in architectures for metropolitan-wide networks~\cite{Lancho_10, Xu_09, Ciurana_14} that lead to an increase of the number of quantum users, capacity and distance reach. In particular, we consider the quantum metropolitan optical network (MON) presented in~\cite{Ciurana_14} as a  starting point for a framework to allow entanglement distribution together with other quantum information technologies.

The goal of this work is to present practical metropolitan optical network designs that support entanglement distribution  between any pair of users, while enabling direct optical paths to allow quantum and classical communication between them. Hence, the network provides all basic transmission resources needed by quantum information technologies. To be practical, it must integrate well with the deployed infrastructure, allowing for shared use, and be based on standard, readily available telecom components designed for mass deployment.

The rest of this paper is organized as follows. First, in Section~\ref{sec:source} we describe the operation mode of the entanglement source considered in our work, and the \textit{canonical} architecture of a metropolitan optical network (MON) in Section~\ref{sec:qkdmetro}. In Section~\ref{sec:entonly} we propose a metropolitan optical network, in which only entangled quantum signals are distributed. Later, in Section~\ref{sec:grid}, we present a channel plan that transmits quantum, one-way and entangled, and conventional signals. Based on this channel plan, we design two networks employing different architectures in Section~\ref{sec:passive} and Section~\ref{sec:active}, respectively. Finally, we discuss both architectures and draw some conclusions in Section~\ref{sec:discussion}.

\section{Broadband Source of Entangled Pairs of Photons}
\label{sec:source}

\begin{figure}
\centering
\includegraphics[width=0.68\linewidth]{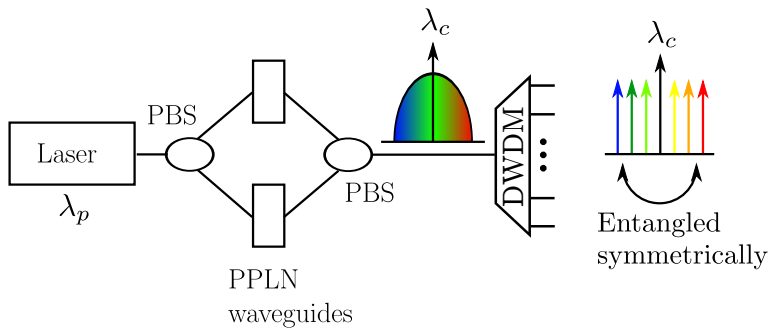}
\caption{Scheme and output of a broadband source of entangled photon-pairs (for details see~\cite{Herbauts_13}). A laser pumps the PPLN waveguides. SPDC generate photon pairs over a broad spectrum that is sliced into DWDM channels using a demultiplexer. Hence, DWDM channels are entangled symmetrically. PBS stands for polarizing beam splitter.}
\label{fig:source}
\end{figure}

The scheme of a broadband source generating entangled photon pairs is shown in Fig.~\ref{fig:source} and its output is based on a spontaneous parametric down conversion process (SPDC). A narrow-band continuous wave (CW) laser diode at $\lambda_p$ pumps the periodically-poled lithium niobate (PPLN) waveguides to obtain degenerate photons with a sprectrum centered at $\lambda_c=2*\lambda_p$. A photon at the lower part of the spectrum is frequency correlated with a one of the upper part and both are polarization entangled each with the other. The wavelength $\lambda_p$ is selected such that the pairs are generated into (the spectral slots of) dense wavelength-division multiplexing (DWDM) channels. Note that the signal produced by the source has constant spectral density, therefore using a denser DWDM grid in order to connect more users is in a trade-off with a lower photon rate per channel. The broadband output makes these sources especially suitable for use in DWDM-based networks, since a single source can serve many users simultaneously.

As a result, photons of each DWDM channel are entangled with the ones in a corresponding other channel. This is of special importance for the network, in which we want to route a couple of entangled DWDM channels from the source to a pair of users. Selecting one of the DWDM channels fixes the second one that has to be used. Note that, when such source is switched on, the entire broadband signal is always produced. We cannot produce entangled photon-pairs only at certain wavelengths within the output spectrum. Thus, in order to control which pairs are distributed, we need to use network components at the output of the source (e.g. switches).

In particular, we consider a source as in~\cite{Herbauts_13} with the following characteristics: (i) $\lambda_p = 775 \pm 5$ nm, (ii) $\lambda_c$ in the C band and (iii) with a spectral width of 70~nm. Based on the same scheme, sources using shorter PPLN waveguides or chirped poling structures~\cite{Ultrabroadband} could generate entangled photon pairs over a spectral width of the entire S-C-L-telecom band. In terms of output power, the source generates $4.5\times10^5$ pairs/s/mW/GHz. For waveguide-based sources typically pumping powers below $1$mW are needed not to be limited by unwanted multi-pair effects~\cite{doublepairs}.

\section{Metropolitan Optical Networks}
\label{sec:qkdmetro}

\begin{figure}
\centering
\includegraphics[width=0.7\linewidth]{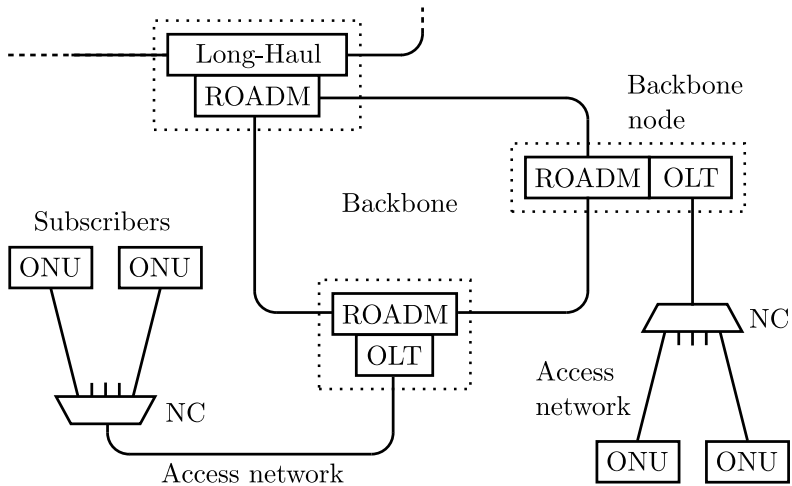}
\caption{Architecture of a \textit{canonical} metropolitan optical network. Users are connected to tree-type access network via a network component. All the access networks are connected to a ring-shaped or mesh backbone network via an optical line terminal and a reconfigurable optical add-drop module.}
\label{fig:mon}
\end{figure}

The architecture of a \textit{canonical} metropolitan optical network~\cite{Ramaswami_09} is depicted in Fig.~\ref{fig:mon}. In the figure, tree-type access networks based on passive optical network (PON) technology are connected to a core or backbone network. This architecture allows users to communicate with other users within the same access network, or in a different access network through the backbone.

End users (i.e., subscribers) in the access networks are connected to optical network units (ONU) which are then connected by a network component (NC). Their signals are multiplexed using time- or wavelength-division multiplexing (TDM or WDM, respectively). In the TDM case, the NC is an optical splitter, while in the WDM case it is a dense WDM (DWDM) multiplexer based on arrayed waveguide grating (AWG) technology. The NC is then connected to an optical line terminal (OLT), located at the premises of the service provider. The OLT coordinates the traffic within the access network and performs the necessary conversions to introduce the signal in the backbone network.

Usually, the backbone network topology is either a ring (as in Fig.~\ref{fig:mon}) or a mesh, and coarse WDM (CWDM) is the most common technology used by carriers to multiplex its traffic. The network consists of multiple backbone nodes that switch signals (depending on their wavelength) between the backbone and the access networks. For this purpose, an optical add-drop multiplexer (OADM)\footnote{Also common in its reconfigurable version, i.e., reconfigurable OADM or ROADM.} is connected to the OLT. In our network, both devices, OADM and OLT, are considered to be situated in the same location, which allows us to propose a backbone node that does not interrupt the quantum signals by performing, for example, electro-optical conversions.

In this work we are looking for simple deployment and maintenance, hence we will stick to the presented typical technologies for the design of quantum metropolitan optical networks. In particular, we propose CWDM~\cite{CWDM_03} for the backbone and DWDM~\cite{DWDM_02} for the access networks. This allows to use common commercial components. Note that in contrast to TDM-based networks, the proposed one does not require any synchronization and allows for simultaneous communications. Hereafter we use \textit{CX} and \textit{DX} to refer to the CWDM channel with $\lambda=X$~nm and the DWDM channel with $\lambda=X$~nm, respectively. Further, whenever required for the calculations, we will consider the 100~GHz DWDM ITU-T G.694.1 grid and 32-channels AWGs. The fiber distances in the network are assumed to be: $1$~km between users and the AWG, $3.5$~km between the AWG and a backbone node, and $4$~km between neighboring backbone nodes. Insertion losses of common network components~\cite{corning, flyin, polatis, Nortel} are shown in Table~\ref{tab:insertion-losses}.

\begin{table}
\centering
\caption{Insertion Losses} 
\label{tab:insertion-losses}
\begin{tabular}{l l l}
\hline\hline
       & Operating &                \\
       & wavelength &                \\
Component & range (nm) & Insertion loss (dB) \\
\hline
Single-mode fibre ITU-T G.652 & 1550 & $0.2$ per km \\
Single-mode fibre ITU-T G.652 & 1310 & $0.32$ per km \\
$1:2$ Splitter & 1260 -- 1610 & $3.6$ \\
$1:32$ Splitter & 1260 -- 1610 & $16.5$ \\
1-channel CWDM OADM & 1270 -- 1610 & $0.4-0.6$ \\
1-channel DWDM OADM & 1525 -- 1610 & $0.4-0.6$ \\
4-channels CWDM mux & 1270 -- 1610 & $1$ \\
1310/1550 WDM mux & 1260 -- 1360 & $0.5$ \\
& \& 1500 -- 1600 & \\
Bandpass filter & & $0.4-0.6$ \\
Circulator & & $0.8$ \\
32-channels AWG (100~GHz) & 1533 -- 1558 & $3$ \\ 
$4 \times 4$ to $192 \times 192$ Switch & 1270 -- 1675 & $1$ \\
\hline
\hline
\end{tabular}
\end{table}

The aim of this work is to design a metropolitan optical network, based on the above mentioned technologies, and a design of the backbone nodes that can enable the routing of entangled pairs of photons between any two network users. In addition to routing, the overall design must take into account other aspects such as the maximum number of users that can be simultaneously connected to the network, as well as the loss budget (that, for instance, directly affects the maximum distance and speed at which two users can be connected).

\section{Entanglement-only Metropolitan Optical Network}
\label{sec:entonly}

We start by proposing a design for metropolitan optical networks supporting only entanglement distribution. Despite its simplicity,this design presents an advance over the state-of-the-art of entanglement-only telecom optical networks~\cite{Brassard_04, Lim_10, Ghalbouni_13, Herbauts_13} and also serves the purpose to illustrate the operation mode of the sources and their integration in WDM-based networks.

\subsection{Single Access Network}

\begin{figure}
\centering
\includegraphics[width=0.75\linewidth]{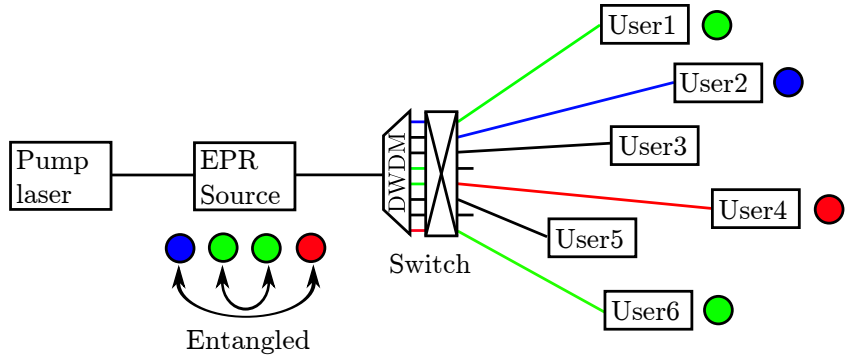}
\caption{Broadband entanglement source connected to the users via a switch, which is necessary to enable all possible user pairings. This is equivalent to a WDM-based access network.}
\label{fig:ent-access}
\end{figure}

The first approach is to directly connect the users to the output of the source. This is shown in Fig~\ref{fig:ent-access}, including an intermediate switch that is required to change among all possible user pairings. The number of input and output channels of the switch must be matched to the available number of DWDM-channels of the source and the number of users, respectively. The resulting point-to-multipoint network is a WDM-based access network.
The number of established connections at a given time is limited by the bandwidth of the source. More users can be connected physically to the switch to receive available photons on demand. With this simple solution only users attached to a central point of a star network could be connected.

\subsection{Two Access Networks}

A more flexible network can be obtained by adding a backbone network to combine the simple access network from above. Thereby the architecture changes due to the shift of the source to the backbone. Let us assume now that all DWDM channels from the access network are contiguous. We can group a set of them within the width of a single ITU-T G.694.2 CWDM channel. Therefore, since the spectrum of the source is wider than a CWDM channel and by choosing the pump wavelength accordingly, we can generate entangled photon-pairs situated in DWDM channels corresponding to two different CWDM channels. If we send different CWDM channels to different access networks, we can have entanglement shared between two access networks.

\begin{figure}
\centering
\includegraphics[width=0.75\linewidth]{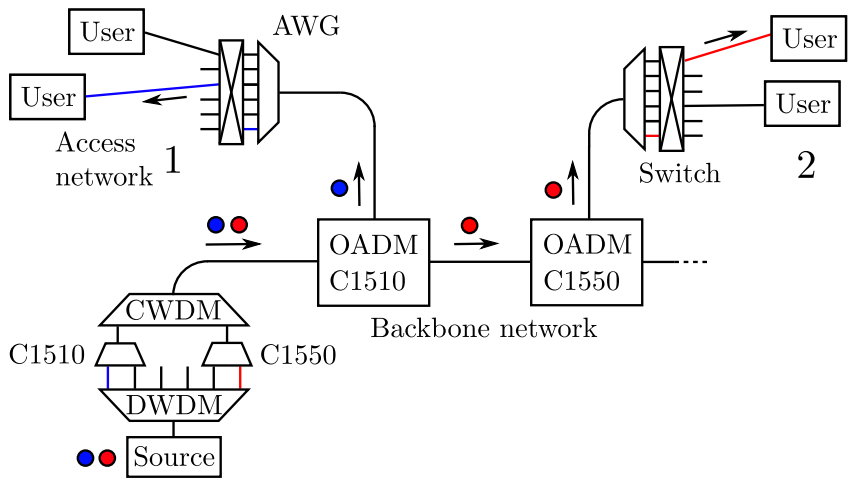}
\caption{Broadband entanglement-source serving two access networks. The output of the source is demultiplexed in DWDM channels and grouped in CWDM channels 1510 and 1550. Each CWDM channel is dropped by an OADM in a different access network.}
\label{fig:ent-metro}
\end{figure}

To do so, at the source, we select the DWDM channels belonging to the chosen CWDM channels (i.e. access networks) and we multiplex them into a single optical fiber. In the optical fiber we add two CWDM OADMs to drop the corresponding CWDM channels at the corresponding access network. The result is a backbone network with an open-ring topology that can be built retaining passive technology. The network is shown in Fig.~\ref{fig:ent-metro}. The source is directly connected to the backbone and distributes entanglement between C1510 and C1550 (blue and red in the figure, respectively). Depending on the particular CWDM channels used, the source configuration will change accordingly by choosing the appropriate multiplexers to group the DWDM channels used.

\subsection{Metropolitan Optical Network}

\begin{figure}
\centering
\includegraphics[width=0.6\linewidth]{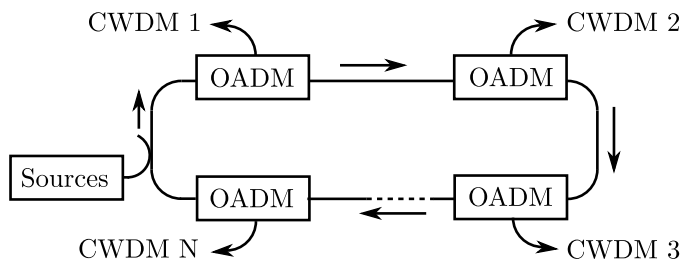}
\caption{Design of a entanglement-only metropolitan optical network with $N$ access networks and a ring-shaped passive backbone. Each access network has assigned a CWDM channel. Sources, connected to the backbone, distribute entanglement over all single CWDM channels and possible pairs, i.e., all single access networks and pairs of them.}
\label{fig:scheme-ent-metro}
\end{figure}

Using these two subnetworks as basic blocks, we can build an entanglement-only metropolitan optical network with $N$ access networks (see Fig.~\ref{fig:scheme-ent-metro}). For example, given $N=3$, we need to address 3 single access networks as in Fig.~\ref{fig:ent-access}, and $\binom{3}{2} = 3$ pairs of access networks as in Fig.~\ref{fig:ent-metro}. Therefore, it requires a maximum of 6 sources: 3 of them to distribute entanglement in each access network, and another 3 to distribute entanglement between the possible 3 pairings of access networks. In this way, any user can share entanglement with any other one. Although the number of sources increases quadratically with $N$ (namely as $\frac{N (N+1)}{2}$, the number of access networks is severely limited by the loss budget and the available spectrum. Moreover, the actual number of sources can be smaller since some of those combinations could be provided by a single source. For instance, a source that distributes entanglement between C1530 and C1570, can distribute it at the same time to the users in a single access network, e.g.  C1550. Regarding the physical allocation of the sources, all of them are grouped as a single source and connected to the backbone.

Subsequently, a CWDM channel is assigned to each access network and backbone nodes route these channels to the different access networks. Finally, once within the access network, the AWG slices the CWDM channel into DWDM channels and a switch routes them to the appropriate user. Therefore, a DWDM channel is assigned to each AWG port in the access network thanks to AWG's periodicity\footnote{To simplify matters, we assume that a CWDM channel equals an AWG's spectrum band. However, this may not be the case and then some DWDM channels will be unusable.}. Note that, when emitting a photon (from a pair) using a particular wavelength, the emitter is actually selecting the target/final access network and the AWG port.

\begin{figure}
\centering
\subfloat[Connection scheme 1.]
{
\includegraphics[width=0.42\linewidth]{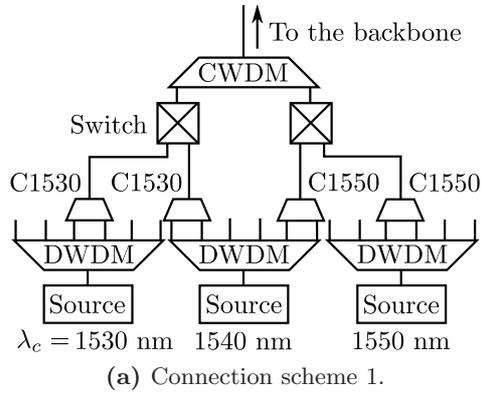}
\label{fig:source-con-1}
}\\
\subfloat[Connection scheme 2.]
{
\includegraphics[width=0.45\linewidth]{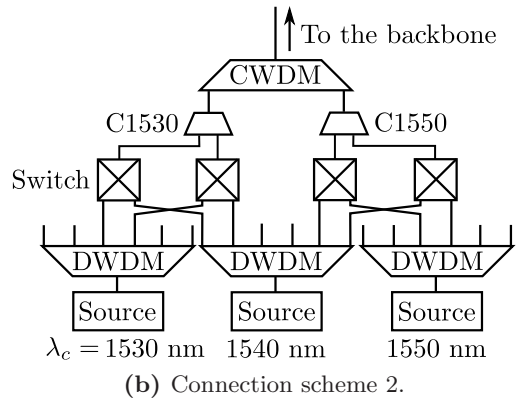}
\label{fig:source-con-2}
}\\
\subfloat[Connection scheme 3.]
{
\includegraphics[width=0.42\linewidth]{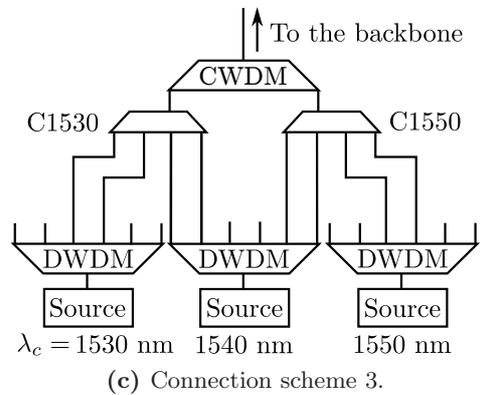}
\label{fig:source-con-3}
}
\caption{Possible connection schemes for the entanglement sources in an entanglement-only metropolitan optical network with 2 access networks. In (a), switches are used to decide which source uses each CWDM channel. In (b), switches decide which source uses each DWDM channel; hence, a CWDM channel is used by multiple sources at the same time. In (c), CWDM channels are also shared among all sources but in a fixed way; a source has assigned always the same DWDM channels.}
\label{fig:source-con}
\end{figure}

However, in this network architecture, multiple sources might try to reach the same access network and thus to use the same CWDM channel. This is is a consequence of the goal to provide entanglement distribution between each pair of users in each access network as well as between any user in this network an any other one in any of the remaining $N-1$ access networks. Hence, we need to carefully connect the sources when grouping them. We propose three solutions (see Fig.~\ref{fig:source-con}). In the first scheme, sources will use the full CWDM channel and a switch will decide which source uses that CWDM channel at each moment. In the second scheme, switches are moved nearer to the sources in order to decide which source uses each DWDM channel of each CWDM channel. Therefore, a CWDM channel is used by multiple sources at the same time. However, this flexibility comes at the cost of using a larger number of switches. In the third scheme, we remove all switches. Instead, we simply distribute the available DWDM channels among the sources. Hence, each source always uses the same DWDM channels.

Now, we calculate the maximum number of users that this network can support using the previously mentioned considerations (see Sec.~\ref{sec:qkdmetro}). First, the maximum number of access networks $N$ is established by the insertion losses. Note that, when using entangled photon-pairs, the losses are the sum of the two paths. Therefore, the worst path in terms of losses in our network is the sum of the two worst paths: from the source to access network $N-1$, and from the source to access network $N$. With a maximal tolerable loss budget of $30$~dB per entaglement based QKD link, we get for $N=8$ access networks: (i) $15.3$~dB from the source to a user in the access network number  $N$, (ii) $14$~dB from the source to a user in the  access network number $N-1$, and (iii) $14+15.3=29.3$~dB overall loss for the QKD link. As a result of the fixed mapping of CWDM channels,  a maximum of 8 CWDM channels are needed. Nevertheless, to reach this upper bound, we need a source able to produce entangled photon-pairs between CWDM 1 and CWDM 8, this is, approx. 160~nm of overall band-width. Hence, depending on the spectral width of the sources used, the actual number of access networks can be lower. Finally, we can estimate the number of users by multiplying $N$ by the number of DWDM channels per CWDM channel (passband of approx. $13$~nm), e.g. $8\cdot\lfloor13/0.8\rfloor=128$ users with a 100~GHz DWDM ITU grid. A higher tolerable loss budget of a QKD link would increase the maximum number of users by allowing more backbone nodes and more users per access network (AWGs with denser DWDM grids).

\section{Adding Direct Optical Paths}
\label{sec:grid}

Up to now, the network has been designed to distribute only entangled photon pairs. Now we will discuss how to incorporate direct communications, quantum and conventional, between users. The former will allow to use a higher variety of quantum information technologies, whereas the latter ones are essential to perform any conventional communication, either required by the quantum information protocols themselves or by external applications.

For this, let us start by adding the conventional signals to the network. As in~\cite{Ciurana_14}, we will use two spectral bands separated by approximately 150~nm: the O band (1260-1360~nm), and the C band and its vicinity (1500-1600~nm). This separation reduces the crosstalk from conventional to quantum signals~\cite{Choi_11} and thus increases the number of simultaneous signals that can be transmitted in the network~\cite{Ciurana_14}. The size of the bands is delimited by their separation and the operating wavelength range of the network components (see Table~\ref{tab:insertion-losses}).

Based on the source characteristics, we assign the O band to the conventional signals and the C band, and its spectral vicinity, to the quantum ones. Each band is then divided into CWDM channels and a pair of them, one in the C band and the other in the corresponding AWG period in the O band, is assigned to each access network. The operation mode of the backbone nodes and the AWGs at the access networks remains the same. Therefore, now each AWG port has assigned two DWDM channels: one containing single photons that are entangled with those in another DWDM channel in the C band, and another one, in the O band, containing conventional optical pulses. Note that both should be separated at the receiver using a filter.

For the connection of the sources, we use the scheme depicted in Fig.~\ref{fig:source-con-3}. Since DWDM channels are assigned and fixed to a specific source, each source is independent of the rest. As long as DWDM channels are distributed over the sources correctly (e.g., a DWDM channel is not assigned to more than one source), sources can be separated and connected at different points of the backbone network. In addition, by assigning and fixing a dedicated source to a subset of CWDM channels, we avoid wavelength-tuning of the source on-the-fly, a complex task that would affect all users receiving from the source.

So far and in contrast to~\cite{Ciurana_14}, CWDM channels are shared among all sources, without leaving room for one-way quantum signals. We could use a third spectral band for one-way quantum signals. However, we are already occupying both low-attenuation telecommunication windows, and to use a third CWDM channel per access network would also increase the losses at the backbone node (as more components would be needed for the OADM). Therefore, we propose to share the quantum band between one-way single photon pulses and entangled pairs. To this end, among the DWDM channels assigned to each source, only a set of them will actually be connected to the source. The rest are left free for one-way quantum signals\footnote{Vice versa, one-way systems must not emit quantum signals at the DWDM channels reserved for entanglement distribution.}. Furthermore, this set of channels is fixed. Instead of reconfiguring the source, we use the switch at the access network to connect the user to the corresponding AWG port.

\begin{figure}
\centering
\includegraphics[width=0.85\linewidth]{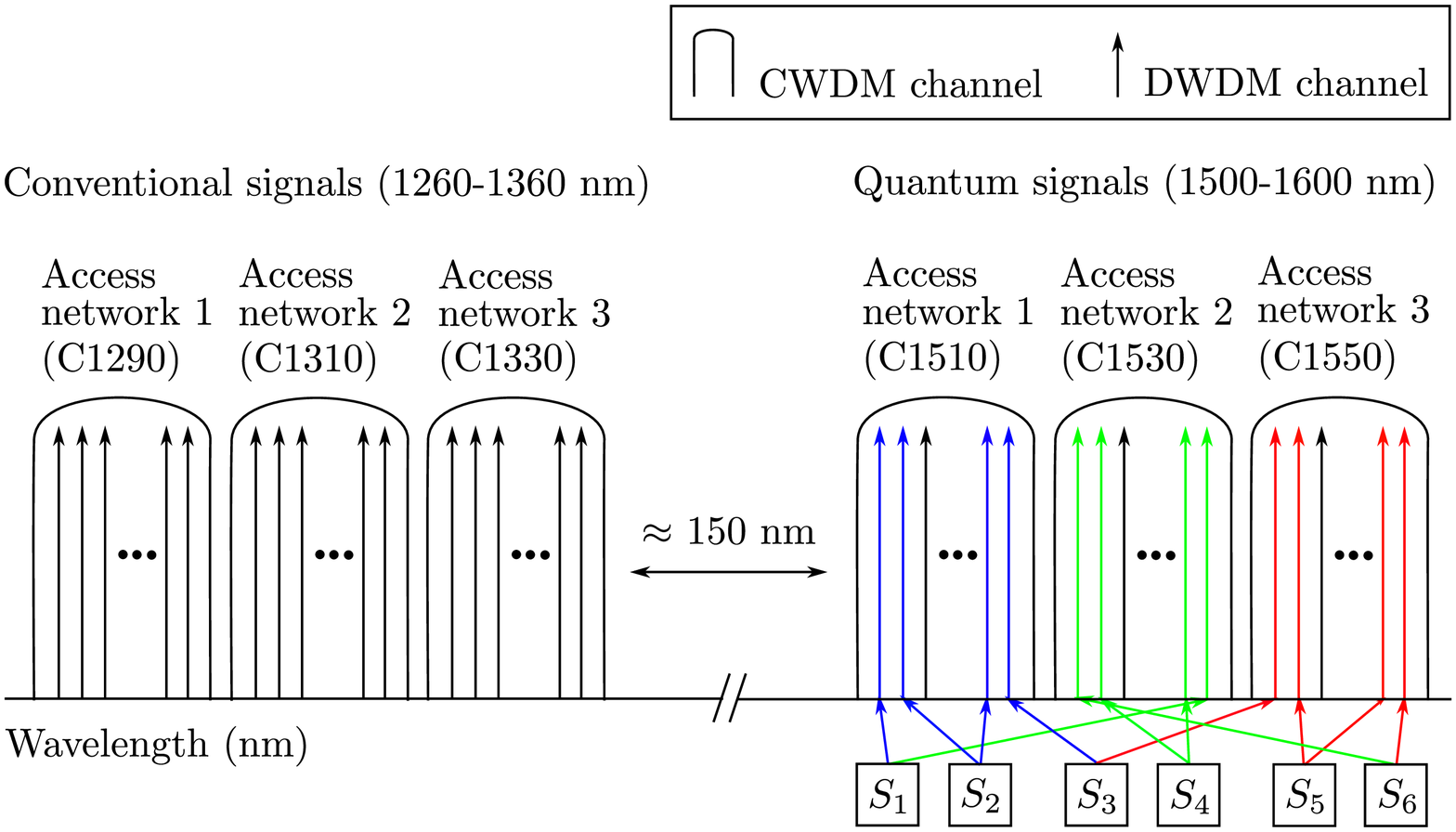}
\caption{Channel plan for a quantum metropolitan optical network. Each access network has assigned two CWDM channels, for quantum and conventional signals. They are spectrally separated to avoid any crosstalk. Within the CWDM channel, DWDM channels are used for one-way communications or entanglement distribution. The figure shows how entanglement sources ($S_x$) are arranged to ensure entanglement among any pair of users in the network.}
\label{fig:spectrum-ent}
\end{figure}

The resulting channel plan is shown in Fig.~\ref{fig:spectrum-ent}. The figure shows the arrangement of bands, CWDM and DWDM channels. Colored arrows indicate DWDM channels used by entanglement sources ($S_x$). In this case, using 6 sources we can distribute entanglement to any pair of users of the network, within the same access or in separate ones. The rest of DWDM channels (in black) are available for one-way signals. Users can connect to each type configuring the switches at the access networks.

A substantial advantage of our design is its scalability when facing a demand increase of entangled channels. We can gradually provide more channels by just connecting the outputs of the source to the DWDM multiplexer. However, this will decrease the available channels for one-way quantum communications.

\section{Quantum Metropolitan Optical Network with Entanglement Distribution}
\label{sec:passive}

Here we modify the previous network design in order to use more sources of entangled photon pairs in the new channel plan. As before, we will stick to passive optical technology for the backbone nodes, which makes them static from the routing point of view, i.e., signals are always routed in the same way.

\subsection{Backbone Node}

A prototype of the backbone node is provided in Fig.~\ref{fig:rigid-node}. It is based on a typical OADM, and is responsible for routing CWDM channels to access networks, and adding the source's signal as well as signals emitted from the access network to the backbone traffic. The operation mode is as follows:

\begin{figure}
\centering
\includegraphics[width=0.45\columnwidth]{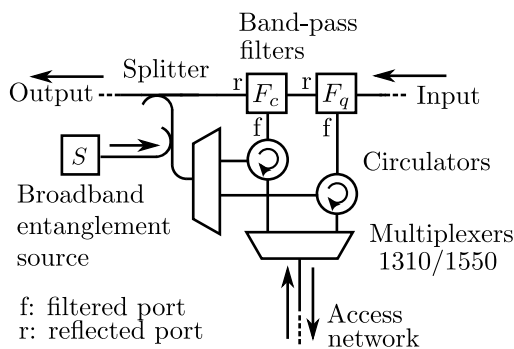}
\caption{Design of a passive backbone node for a ring-shaped backbone that includes a broadband source of entangled photon-pairs.}
\label{fig:rigid-node}
\end{figure}

\subsubsection{Drop}

A signal reaches the OADM through the input port. Then, a pair of quantum and conventional CWDM channels are filtered via two band-pass filters, $F_q$ and $F_c$, respectively, centered at the appropriate wavelength. Both are routed using two circulators and later multiplexed by a 1310/1550 WDM mux. Finally, the multiplexed signal leaves the backbone node through the add/drop (access network) port.

\subsubsection{Add}

A signal reaches the OADM through the add/drop port (access network). It is then demultiplexed into quantum and conventional bands by a 1310/1550 WDM mux. Both bands are routed using two circulators to another 1310/1550 WDM mux that joins them again. At this point, a couple of $1 \times 2$ splitters combines this signal with the output by the entanglement source ($S$), and with the signal reflected by the band-pass filters $F_q$ and $F_c$. The combined signal leaves the OADM through the output port. The signal output by the OADM has then three origins: backbone, access network, and entanglement source.

Note that, if the backbone node does not have a source $S$, one of the splitters can be removed.

Table~\ref{tab:losses-oadm-ent} shows the losses of the node, calculated using the values from Table~\ref{tab:insertion-losses}. Clearly, the splitter is the component that has a larger impact on the overall losses. Nevertheless, we put the splitter from the source in the \textit{add path} to minimize its effect: pass trough and drop signals are not affected. In a communication, a signal will only suffer one time these extra losses, no matter how far the destination is. Likewise, we use an unbalanced splitter not to hinder the signals coming from the access network (only $+0.8$~dB for a 90:10 splitter). Meanwhile, the increment of losses at the source can be counteracted by increasing the pumping power.

\begin{table}
\caption{Losses of the passive backbone node with entanglement-capability depicted in Fig.~\ref{fig:rigid-node}.}
\label{tab:losses-oadm-ent}
\centering
\begin{tabular}{l l l l}
\hline\hline
Action & Losses Conv. & Losses Quantum & Losses Ent. \\
\hline
Add & $6.2$~dB & $6.2$~dB & $3.6$~dB  \\ 
Pass & $4.8$~dB & $4.8$~dB & $4.8$~dB \\ 
Drop & $2.3$~dB & $1.7$~dB & $1.7$~dB \\
\hline\hline
\end{tabular}
\end{table}

\subsection{Access network}

The current configuration of the access network routes all upstream signals to the backbone network. If two users from the same access network want to communicate directly, their signals will cross the entire backbone before reaching the receiver. This inefficient operation mode can be decisively improved by creating a short path within the access network. We propose to use a larger switch than needed and to use the extra ports to create loops in the network's side of the switch, i.e., \textit{return paths}. Therefore, two users can connect to the user's side of the switch and use the loop to communicate directly. This is a simple, cost-free and local solution that does not introduce extra losses or modify the channel plan.

\subsection{Network Design}

\begin{figure*}
\centering
\includegraphics[angle=90,width=0.5\textwidth]{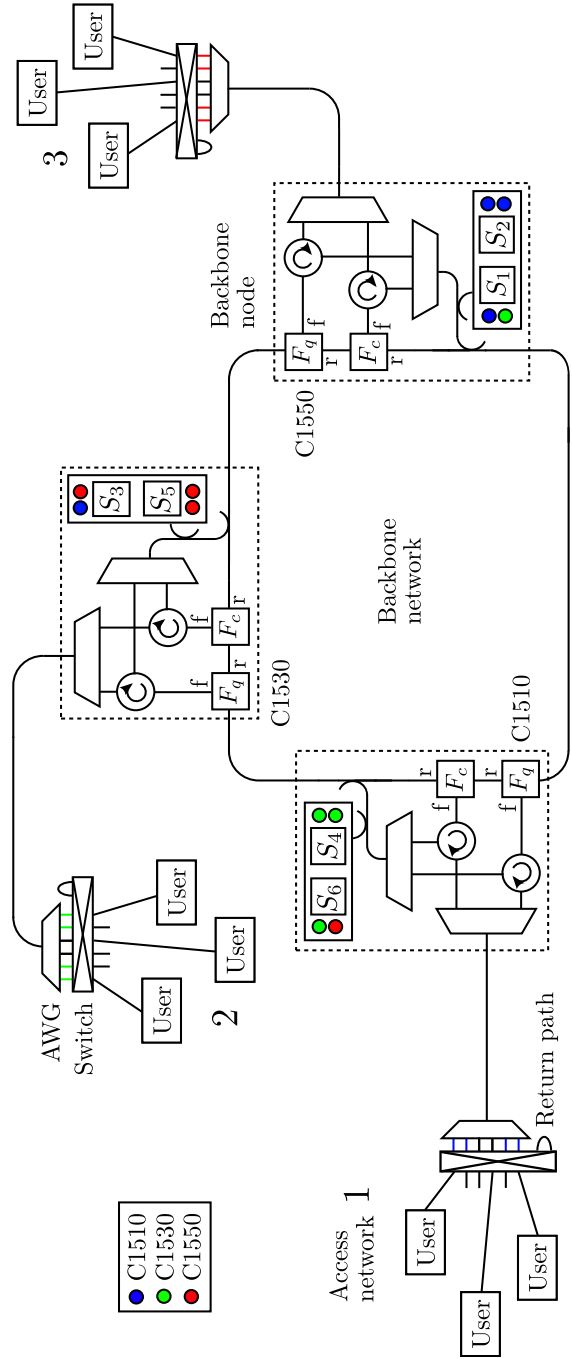}
\caption{Quantum metropolitan optical network based on the design shown in Section~\ref{sec:qkdmetro}. Besides allowing one-way communications, quantum and classical, the network is also capable of distributing entangled-photon pairs among any pair of users of the network. The pairs of circles represent the entangled photon-pairs produced by each source and the color indicates the CWDM channel of the photon.}
\label{fig:scheme-rigid}
\end{figure*}

As an example, we build a quantum metropolitan optical network using the described modifications. The network, depicted in Fig.~\ref{fig:scheme-rigid}, has three access networks ($A_x$) and a backbone ring. The conventional and quantum CWDM channels assigned to each access networks are: (C1290, C1510) for $A_1$, (C1310, C1530) for $A_2$, and (C1330, C1550) for $A_3$. The entanglement sources ($S_x$) are configured as in Fig.~\ref{fig:spectrum-ent}:
\begin{itemize}
\item $S_1$ serves $A_1$ (C1510) and $A_2$ (C1530) with $\lambda_c=1520$~nm
\item $S_2$ serves only $A_1$ (C1510) with $\lambda_c=1510$~nm
\item $S_3$ serves $A_1$ (C1510) and $A_3$ (1550) with $\lambda_c=1530$~nm
\item $S_4$ serves only $A_2$ (C1530) with $\lambda_c=1530$~nm
\item $S_5$ serves only $A_3$ (C1550) with $\lambda_c=1550$~nm
\item $S_6$ serves $A_2$ (C1530) and $A_3$ (C1550) with $\lambda_c=1540$~nm. 
\end{itemize}
This arrangement is represented in the figure using colored circles located near each source. They represent the entangled photon-pairs generated by the source. The color indicates the CWDM channel of the photon (i.e., the destination): blue for C1510, green for C1530, and red for C1550. Note that the sources are deployed in a way that they always distribute photon-pairs among the next and second next access networks in order to use the shortest paths.

We calculate the path losses in Table~\ref{tab:losses-oadm-path} using the considerations from Section~\ref{sec:qkdmetro}, and values from Table~\ref{tab:insertion-losses} and Table~\ref{tab:losses-oadm-ent}. We use the notation $x$-closest to denote proximity: 0-closest would be the same access network, 1-closest the immediate next in the backbone direction, 2-closest the immediate next after the 1-closest, etc. Note that, in this design, there is no short-path between an entanglement source and the immediate access network. Therefore, in order to reach the 0-closest access network, the signal has to travel the entire backbone network. For instance, in Fig.~\ref{fig:scheme-rigid}, the 3-closest access network is actually the 0-closest, since, after crossing two backbone nodes, you are back to the immediate access network.

As shown, with a tolerable loss budget of $30$~dB, our network design allows one-way quantum communications between non-neighboring access networks, e.g., from $A_1$ to $A_3$. For entangled communications, the longest path is between a neighboring access network and the next one, e.g., from $S_2$ to $A_1$ and $A_2$. The result is slightly worse in this case because the source always communicates with farther access networks. In the case of one-way, the source is already in one of the ends.

\begin{table}
\caption{Path losses from an emitter (user or source) in a QKD-MON with a fixed-ring backbone (Fig.~\ref{fig:scheme-rigid}). Values calculated using Table~\ref{tab:insertion-losses} and Table~\ref{tab:losses-oadm-ent}.} \label{tab:losses-oadm-path}
\centering
\begin{tabular}{l l l l}
\hline\hline
Path to & Losses Conv. & Losses Quant. & Losses Ent. \\
\hline
0-closest access network & $2.64$~dB & $2.4$~dB & -   \\
1-closest access network & $20.66$~dB & $18.5$~dB & $11$~dB   \\
2-closest access network & $26.74$~dB & $24.1$~dB & $16.6$~dB \\
3-closest access network & $32.82$~dB & $29.7$~dB & $22.2$~dB \\
\hline\hline
\end{tabular}
\end{table}

We estimate the maximum number of users\footnote{Here we refer to the maximum number of users that could in principle communicate simultaneously. In practice, this is limited by the noise produced by conventional signals.} following the same procedure as before. The first limitation comes from the losses. With a tolerable loss budget of $30$~dB, the network design is limited to 3 access networks (approx., 48 users). Adding a fourth access network would inevitably require a source to distribute pairs among non-neighboring access networks (e.g., $A_2$ and $A_4$) and to have one-way quantum communications between 3-closest access networks. Beyond the losses, we face again the spectral width of the source and then the number of CWDM channels available.

\section{Network Design Based on a Mesh-Shaped Active Backbone}
\label{sec:active}

The network proposed in the previous section fulfils the intended goals, but falls short when facing a considerable increase in the number of users. The objective now is to devise a larger quantum metropolitan optical network using current quantum technology. For this, we conserve the discussed channel plan and access networks, but modify the backbone architecture using reconfigurable nodes and a mesh topology in order to reduce the losses and ensure architecture flexibility.

\subsection{Backbone Node} 

\begin{figure}
\centering
\includegraphics[width=0.5\columnwidth]{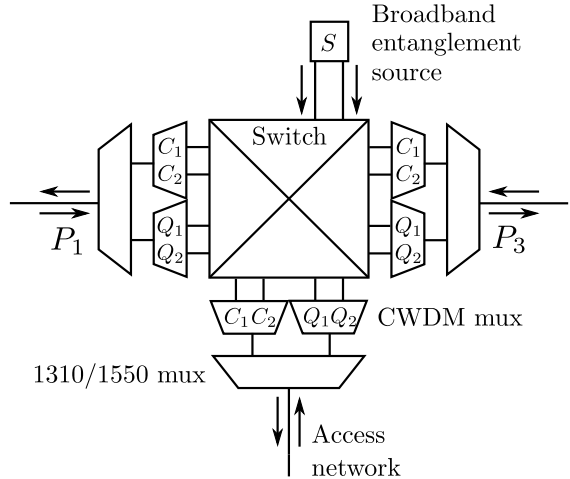}
\caption{Design of an active backbone node for a mesh-based backbone that includes a broadband source of entangled photon-pairs. Incoming signals are demultiplexed into quantum and conventional bands, and then into CWDM channels. These are routed to their corresponding port via a switch. All signals are multiplexed again before leaving the node.}
\label{fig:switched-node}
\end{figure}

An active and reconfigurable version of the backbone node is shown in Fig.~\ref{fig:switched-node}. The design is a transparent optical cross-connect, also referred to as photonic cross-connect (PXC), adapted to our needs. The PXC allows to route any input signal to any output port (even the original one). This opens the door to use a dynamic assignment of CWDM channels for the access networks. The operation mode is simplified to only one function: 

\subsubsection{Cross}

Signals reach the PXC through a port and they are separated into quantum and conventional bands by a 1310/1550 WDM mux. Both bands are demultiplexed into CWDM channels and sent to the switch. The switch routes each channel to the corresponding CWDM mux. CWDM channels are again multiplexed and combined into one signal that leaves the PXC.

In case entanglement sources are needed, they are directly connected to the switch.

Note that due to the new topology, the backbone node uses the wavelength of the signal to decide through which output port the signal goes. Previously, the wavelength was used to decide whether to drop part of the signal or not. Hence, CWDM channels are now assigned to the communication paths instead of to the receivers.

In comparison with the OADM, the losses of the PXC are slightly lower due to the absence of splitters (see Table~\ref{tab:losses-pxc-ent}). However, this depends on the number of CWDM channels used per band. In our calculations, we have considered 4, which is almost the worst-case scenario due to the width of the quantum and conventional bands (approx. 100~nm). Although the figure shows a big switch for all signals, we can use separate switches for each of them without increasing the losses.

\begin{table}
\caption{Losses of the active backbone node with entanglement-capability depicted in Fig.~\ref{fig:switched-node}.} \label{tab:losses-pxc-ent}
\centering
\begin{tabular}{l l l l}
\hline\hline
Action & Losses Conv. & Losses Quantum & Losses Ent. \\
\hline
Cross & $4$~dB & $4$~dB & $2.5$~dB \\
\hline\hline
\end{tabular}
\end{table}

We highlight the fact that the losses of the node are independent of its degree (number of ports). Signals will still cross the same number of components. This is really helpful when dealing with dense areas like metropolitan ones. For example, we can create redundant paths between nodes for resiliency. Another interesting use case is to connect more than one access network per node. We increase the number of users per area but without adding more backbone nodes.

\subsection{Network design}

\begin{figure*}
\centering
\includegraphics[angle=90,width=0.7\textwidth]{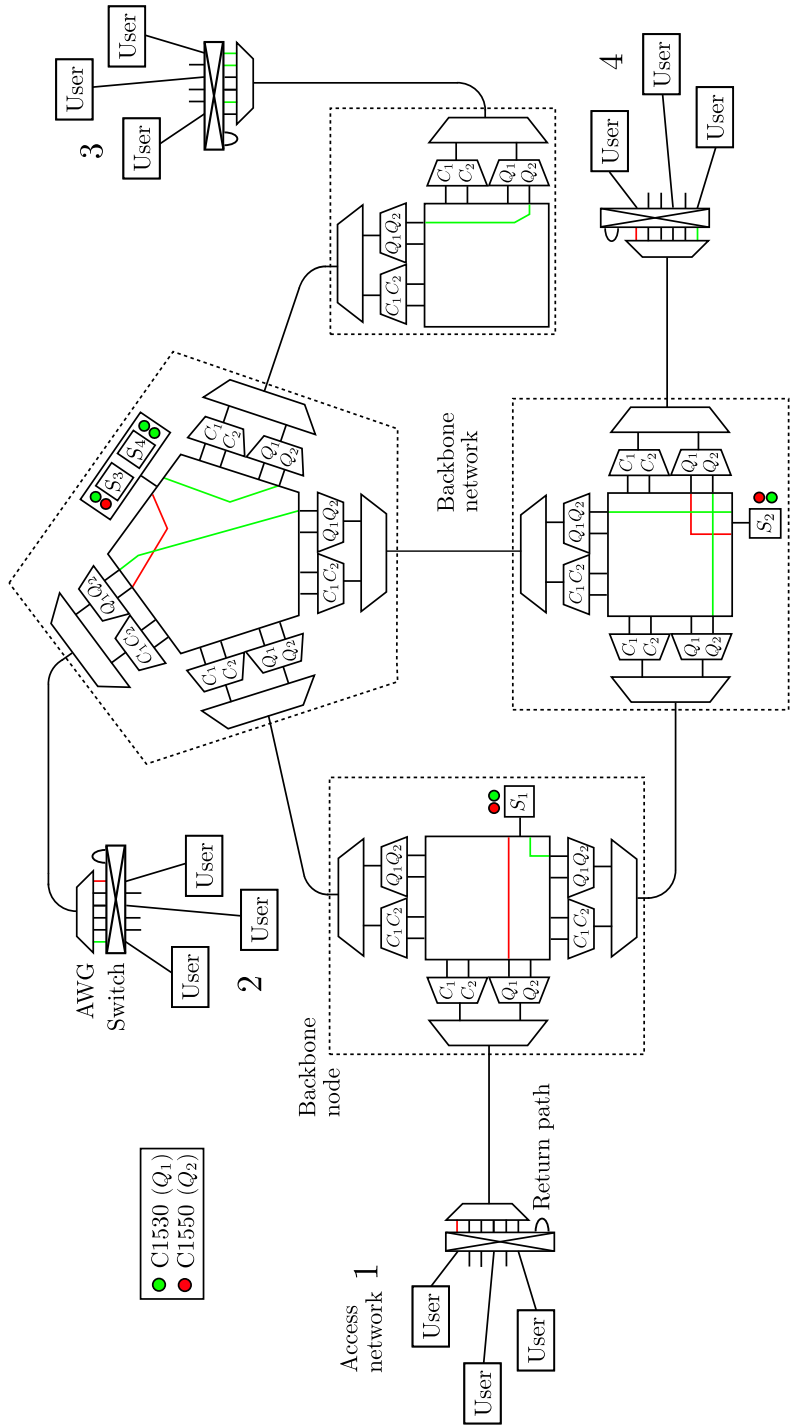}
\caption{Quantum metropolitan optical network with a mesh-type backbone, backbone nodes based on active technology, and 4 access networks ($A_x$). Reconfiguring the nodes allows to interconnect all users using only 2 CWDM channels for quantum signals, but not at the same time. In the depicted configuration, entanglement is shared between $A_1$ and $A_4$, $A_2$ and $A_4$, $A_2$ and $A_3$, and within $A_3$. One-way, quantum and classical communication paths are not configured.}
\label{fig:scheme-switched}
\end{figure*}

Fig.~\ref{fig:scheme-switched} depicts a quantum metropolitan optical network based on active backbone nodes, a mesh topology, and 4 access networks ($A_1$, $A_2$, $A_3$, $A_4$). Even though the network has more users and access networks, it only uses 4 sources and 2 CWDM channels per band: C1290 and C1310, and C1530 and C1550, for conventional and quantum signals, respectively. The communication scheme is still any-to-any, but not simultaneously. CWDM channels are not mapped to any access network. Furthermore, using less CWDM channels allows to reduce the number of different sources. Instead of having different sources, we configure the PXCs to route their signal to different access networks.

In contrast to the previous network design, the backbone nodes have to be configured carefully. We have to enable direct optical paths and entanglement-distribution between any pair of users using a fixed number of CWDM channels in the minimum number of steps. For instance, considering the networks in the figure, we need to create the following optical paths a total of 12 pairs:
\begin{itemize}
\item Enable direct optical paths between: ($A_1$, $A_2$), ($A_1$, $A_3$), ($A_1$, $A_4$), ($A_2$, $A_3$), ($A_2$, $A_4$) and ($A_3$, $A_4$). These are bidirectional, thus ($A_1$, $A_4$) is physically equal to ($A_4$, $A_1$). Moreover, we do not need to address direct paths between users from the same access network since that is already solved by the return paths at the switch.
\item Distribute entanglement between: ($A_1$, $A_1$), ($A_1$, $A_2$), ($A_1$, $A_3$), ($A_1$, $A_4$), ($A_2$, $A_2$), ($A_2$, $A_3$), ($A_2$, $A_4$), ($A_3$, $A_3$), ($A_3$, $A_4$) and ($A_4$, $A_4$), covering thus any pair of users.
\end{itemize}

\begin{figure}
\centering
\subfloat[Configuration 1.]
{
\includegraphics[width=0.55\linewidth]{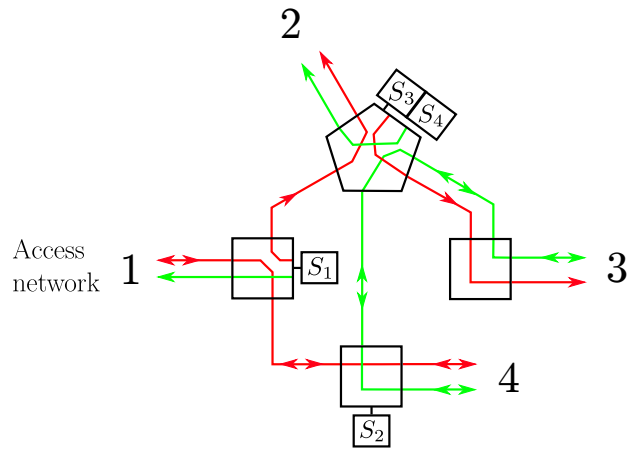}
\label{fig:an-conf1}
}
\\
\subfloat[Configuration 2.]
{
\includegraphics[width=0.55\linewidth]{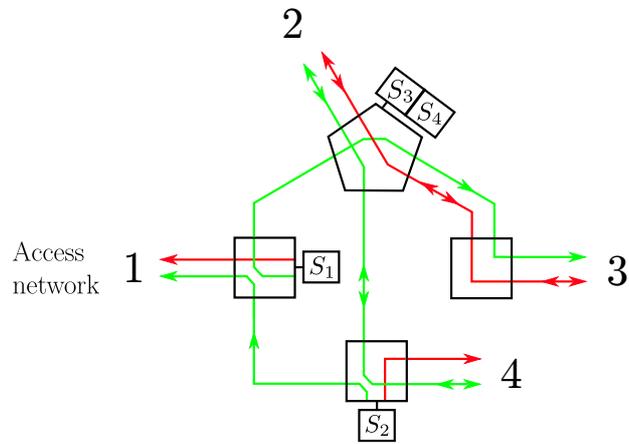}
\label{fig:an-conf2}
}
\\
\subfloat[Configuration 3.]
{
\includegraphics[width=0.55\linewidth]{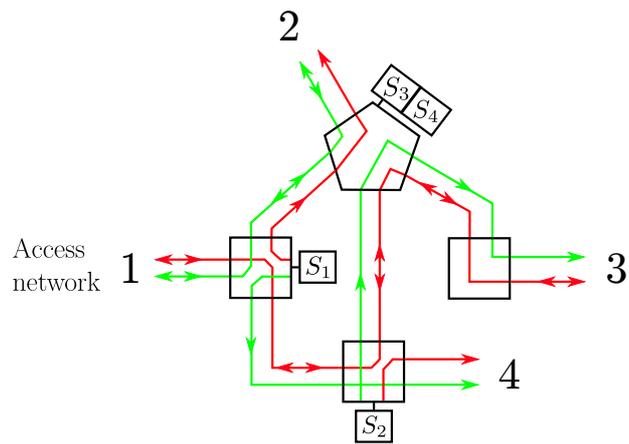}
\label{fig:an-conf3}
}
\caption{Possible node configurations of the quantum metropolitan optical network shown in Fig.~\ref{fig:scheme-switched}. Each backbone node is depicted as an schematic switch-box, and each color represents a CWDM channel for quantum signals. The three configurations cover all communication paths between pairs of access networks.}
\label{fig:an-conf}
\end{figure}

The problem is rather trivial as long as we have enough sources and loss budget. As an example, we show a series of network configurations in Fig.~\ref{fig:an-conf} for the network depicted in Fig.~\ref{fig:scheme-switched}. The figure details each configuration over a simplified version of the network. The following three configurations cover all possible communications between pairs of access networks:
\begin{itemize}
\item Configuration 1: entanglement distribution to the pairs ($A_1$, $A_2$) and ($A_2$, $A_3$), and direct optical paths between the pairs ($A_1$, $A_4$) and ($A_3$, $A_4$). 
\item Configuration 2: entanglement distribution to ($A_1$, $A_3$) and ($A_1$, $A_4$), and direct optical paths between ($A_2$, $A_4$) and ($A_2$, $A_3$).
\item Configuration 3: entanglement distribution to ($A_2$, $A_4$) and ($A_3$, $A_4$), and direct optical paths between ($A_1$, $A_2$) and ($A_1$, $A_3$). 
\end{itemize}
Similarly, the network can be configured also to distribute entanglement to users from the same access network. Then, we can choose to use a fixed set of configurations that cover all possible communication paths and just reuse them over time, or actively switch the configurations depending on the user traffic at the moment.

\begin{table}
\caption{Path losses from an emitter (user or source) in a QKD-MON with a reconfigurable-mesh backbone (Fig.~\ref{fig:scheme-switched}). Values calculated using Table~\ref{tab:insertion-losses} and Table~\ref{tab:losses-pxc-ent}.} \label{tab:losses-pxc-path}
\centering
\begin{tabular}{l l l l}
\hline\hline
Path to & Losses Conv. & Losses Quant. & Losses Ent. \\
\hline
0-closest access network & $2.64$~dB & $2.4$~dB & $7.4$~dB \\
1-closest access network & $20.16$~dB & $18.6$~dB & $12.2$~dB \\
2-closest access network & $25.44$~dB & $23.4$~dB & $17$~dB   \\
3-closest access network & $30.72$~dB & $28.2$~dB & $21.8$~dB \\
\hline\hline
\end{tabular}
\end{table}

We recalculate the path losses in Table~\ref{tab:losses-pxc-path}. As expected, the values are lower because the backbone node introduces less losses. Moreover, entanglement sources can now communicate with receivers located at the immediate access network, i.e., the access network that is connected to the same backbone node as the source. Keeping in mind the $30$~dB tolerable loss budget for QKD, the new design is limited to quantum communications between access networks separated by two intermediate backbone nodes: one way, $28.2$~dB, and entangled, $29.2$~dB ($12.2+17$).

Besides allowing farther communications, the number of users is also no longer a problem. In terms of loss budget, the mesh topology allows to add access networks without surpassing the worst-case path (in terms of losses) of the network by adding links that bypass intermediate backbone nodes. On the other hand, using PXCs and a dynamic assignment of CWDM channels allows to rotate the assignment of CWDM channels if we have more access networks than CWDM channels. Hence, the spectrum width of the source is not a limitation any more. Finally, even if all available CWDM channels are used, the mix of active nodes plus mesh topology allows to reuse them in different parts of the network at the same time (see Fig.~\ref{fig:an-conf}).

In the end, the network can grow and the only effect will be an increase in the number of configurations needed to cover all possible communication paths. It is equivalent to decomposing the reconfigurable network into smaller fixed ones.

\section{Discussion}
\label{sec:discussion}

In this paper we have proposed several optical network designs that allow the distribution of entangled photon-pairs alongside with one-way quantum and conventional communications. This is especially helpful if we want to embrace as many quantum information technologies as possible and also to share the costs of the infrastructure.

Starting from a single access network, first we have shown a simple metropolitan optical network dedicated only to the distribution of entangled photon-pairs. It requires few components, which reduces the costs and the losses. The network is based on a mixture of coarse and dense wavelength multiplexing and can serve up to 8 access networks (approx. 128 users using a 100~GHz DWDM grid). Nevertheless, entanglement-based protocols and other quantum information technologies typically require also a direct optical path for quantum and conventional communications. For this task, we have proposed a channel plan that includes all these signals and two different network designs.

The first complete network design that includes all the signals is based on a backbone with ring topology and fixed passive nodes (OADMs). It is limited to 3 access networks with current quantum technology (approx. 48 users). In order to go beyond this network size, we have explored a more flexible network architecture based on a backbone with mesh topology and reconfigurable active nodes (PXCs).

The new design is not limited in terms of users. Moreover, with the same tolerable loss budget, it permits farther communications. The flexibility of using at will different paths to reach the same destination also makes the network resilient to attacks and link failures although the passive components used in the previous design are more robust and reliable.  Nevertheless, these benefits come at a price. First, both the deployment and operational costs increase considerably. Switches are not, in general, a cheap component (depending on the number of ports) and active nodes require conditioned facilities, a constant supply of energy, more maintenance and a management protocol. Moreover, the management layer becomes more complex as we add nodes that, in the end, can compromise the scalability. Adding or removing backbone nodes, links and/or access networks in this case requires a complete reconfiguration.

\section*{Acknowledgment}

This work has been partially supported by the project HyQuNet, TEC2012-35673, funded by \textit{Ministerio de Econom\'\i a y Competitividad}, Spain, and by the project QKD-Telco (www.qkd-telco.at), funded by the \textit{Austrian Research Promotion Agency} (FFG) under contract 835926. 

\bibliographystyle{unsrtnat}
\bibliography{entangled-metro}

\end{document}